\documentclass[aps,preprintnumbers,amsmath,amssymb,showpacs,nofootinbib,twocolumn,superscriptaddress]{revtex4}
\usepackage{amsfonts}
\usepackage{mathrsfs}
\usepackage{graphicx}
\usepackage{amsmath}
\usepackage{amssymb}
\usepackage{bm}
\usepackage{bbm}
\usepackage{epsfig}
\usepackage{multirow}
\usepackage{float}
\usepackage{array}
\makeatletter
\def\hlinew#1{%
  \noalign{\ifnum0=`}\fi\hrule \@height #1 \futurelet
   \reserved@a\@xhline}
\usepackage{array}
\newcommand{\PreserveBackslash}[1]{\let\temp=\\#1\let\\=\temp}
\newcolumntype{C}[1]{>{\PreserveBackslash\centering}p{#1}}
\newcolumntype{R}[1]{>{\PreserveBackslash\raggedleft}p{#1}}
\newcolumntype{L}[1]{>{\PreserveBackslash\raggedright}p{#1}}

\def\bfalpha{\mbox{\boldmath $\alpha$}}
\def\bfpsi{\mbox{\boldmath $\psi$}}
\def\bfeta{\mbox{\boldmath $\eta$}}

\def\OMIT#1{}

\newcommand{\nn}{\nonumber}

\newcommand{\beq}{\begin{equation}}
\newcommand{\eeq}{\end{equation}}
\newcommand{\bqa}{\begin{eqnarray}}
\newcommand{\eqa}{\end{eqnarray}}

\begin{document}

%%%%%%%%%%%%%%%%%%%%%%%%%%%%%%%%%%%%%%%%%%%%%%%%%%%%%%%%%%%%%%%%%%%%%%%%%%%%%%
\title{\mbox{}\\[11pt]
$\bm{\mathcal O}\bm{(}\bfalpha_{\bm s} {\bm v}^{\bm 2}\bm{)}$
correction to $\bm{e}^{\bm +} \bm{e}^{\bm
-}\bm{\to}\bm{J}\bm{/}\bfpsi\bm{+}\bfeta_{\bm{c}}$ at $\bm{B}$
factories}

%%%%%%%%%%%%%%%%%%%%%%%%%%%%%%%%%%%%%%%%%%%%%%%%%%%%%%%%%%%%%%%%%%%%%%%%%%%%%%

\author{Hai-Rong Dong\footnote{donghr@ihep.ac.cn}}
\affiliation{Institute of High Energy Physics, Chinese Academy of
Sciences, Beijing 100049, China\vspace{0.2cm}}

\author{Feng Feng\footnote{fengf@ihep.ac.cn}}
\affiliation{Center for High Energy Physics, Peking University,
Beijing 100871, China\vspace{0.2cm}}

\author{Yu Jia\footnote{jiay@ihep.ac.cn}}
\affiliation{Institute of High Energy Physics, Chinese Academy of
Sciences, Beijing 100049, China\vspace{0.2cm}}
\affiliation{Theoretical Physics Center for Science Facilities,
Chinese Academy of Sciences, Beijing 100049, China\vspace{0.2cm}}

\date{\today}
%%%%%%%%%%%%%%%%%%%%%%%%%%%%%%%%%%%%%%%%%%%%%%%%%%%%%%%%%%%%%%%%%%%%%%%%%%%%%%
\begin{abstract}

We investigate the ${\mathcal O}(\alpha_s v^2)$ correction to the
$e^+e^-\to J/\psi+\eta_c$ process in the nonrelativistic QCD (NRQCD)
factorization approach. Within some reasonable choices of the
relative order-$v^2$ NRQCD matrix elements, we find that including
this new ingredient of correction only mildly enhances the existing
NRQCD predictions. We have also deduced the asymptotic expressions
for the ${\mathcal O}(\alpha_s v^2)$ short-distance coefficients,
and reconfirm the early speculation that at next-to-leading order in
$\alpha_s$, the double logarithm of type $\ln^2 (s/m_c^2)$ appearing
in various NRQCD short-distance coefficients is always associated
with the helicity-suppressed channels.

\end{abstract}

%%%%%%%%%%%%%%%%%%%%%%%%%%%%%%%%%%%%%%%%%%%%%%%%%%%%%%%%%%%%%%%%%%%%%%%%%%%%%%
\pacs{\it 12.38.Bx, 13.66.Bc, 14.40.Pq}

% 12.38.Bx  Perturbative calculations
% 13.66.Bc Hadron production in e-e+ interactions
% 14.40.Pq Heavy quarkonia

\maketitle

%----------------------------------------------------------------------
\section{Introduction}
\label{Introduction}
%----------------------------------------------------------------------

One of the most intensively studied hard exclusive reactions in
recent years is perhaps the double-charmonium production process
$e^+e^-\to J/\psi+\eta_c$ at the $B$ factory energy $\sqrt{s}=10.58$
GeV~\cite{Brambilla:2010cs}. It was initially measured by the
\textsc{Belle} experiment in 2002 with $\sigma[e^+e^-\to
J/\psi+\eta_c]\times{\mathcal B}_{\ge 4}=33_{-6}^{+7}\pm 9$
fb~\cite{Abe:2002rb}, where ${\mathcal B}_{\ge 4}$ is the branching
ratio of $\eta_c$ into 4 or more charged tracks. The first
theoretical
predictions~\cite{Braaten:2002fi,Liu:2002wq,Hagiwara:2003cw}, built
on the lowest-order (LO) calculation in the nonrelativistic QCD
(NRQCD) factorization approach~\cite{Bodwin:1994jh}, were scattered
in the range $2.3\!-\!5.5$ fb, almost one order of magnitude smaller
than the \textsc{Belle} data. Later \textsc{Belle} Collaboration
refined their measurement and gave $\sigma[J/\psi+\eta_c] \times
{\mathcal B}_{>2} = 25.6\pm 2.8\pm3.4$ fb~\cite{Abe:2004ww}, where
${\mathcal B}_{>2}$ denotes the branching fraction for the $\eta_c$
into more than two charged tracks. In 2005, \textsc{BaBar}
Collaboration also measured the same observable and obtained
$17.6\pm2.8^{+1.5}_{-2.1}$ fb~\cite{Aubert:2005tj}.

The disquieting discrepancy between experiment and the LO NRQCD
predictions has spurred a great amount of theoretical endeavors in
the following years. Roughly speaking, most works can be divided
into two major categories, either based on the light-cone
factorization~\cite{Ma:2004qf,Bondar:2004sv,Braguta:2005kr,Bodwin:2006dm},
or based on the NRQCD factorization
approach~\cite{Zhang:2005cha,Gong:2007db,He:2007te,Bodwin:2007ga}
(for the investigations from other theoretical approaches, see
Refs.~\cite{Ebert:2008kj,Guo:2008cf,Zhang:2008ab,Sun:2009zk}).

One crucial step toward alleviating the tension between data and
theory is the discovery of the positive and significant ${\mathcal
O}(\alpha_s)$ correction to the $e^+e^-\to J/\psi+\eta_c$
process~\cite{Zhang:2005cha,Gong:2007db}. This next-to-leading order
(NLO) perturbative calculation was performed in the NRQCD
factorization framework (a similarly large positive ${\mathcal
O}(\alpha_s)$ correction has also later been found for the
$e^+e^-\to J/\psi+\chi_{c0}$
process~\cite{Wang:2011qg,Dong:2011fb}). In contrast, due to some
long-standing theoretical difficulty inherent to the
helicity-suppressed process, by far no one has successfully
conducted the corresponding ${\mathcal O}(\alpha_s)$ correction to
this process in the light-cone approach. In a sense, for the
$e^+e^-\to J/\psi+\eta_c$ process, NRQCD approach seems more
systematic and maneuverable than the light-cone approach.

The relative ${\mathcal O}(v^2)$ correction to $e^+e^-\to
J/\psi+\eta_c$ has also been
addressed~\cite{Braaten:2002fi,He:2007te,Bodwin:2007ga}, where $v$
denotes the typical velocity of the $c$ quark in a charmonium.
Notwithstanding the large uncertainty inherent to relativistic
correction, it was believed that~\cite{He:2007te,Bodwin:2007ga},
including both NLO perturbative and (a partial resummation of )
relativistic corrections, one may achieve the reasonable agreement,
albeit with large uncertainties, between the NRQCD prediction and
$B$ factory data.

The goal of this work is to address the ${\mathcal O}(\alpha_s v^2)$
correction to $e^+e^-\to J/\psi+\eta_c$ in the NRQCD factorization
approach. It is curious to examine its phenomenological impact. On
the other hand, thus far there are only very few basic quarkonium
decay processes whose ${\mathcal O}(\alpha_s v^2)$ corrections have
been calculated, {\it e.g.} $J/\psi\to e^+e^-$~\cite{Luke:1997ys},
and $\eta_c\to\gamma\gamma$~\cite{Jia:2011ah,Guo:2011tz}. Therefore,
it is also theoretically interesting to know the ${\mathcal
O}(\alpha_s v^2)$ effect for the exclusive quarkonium production
process in the first time.

The rest of the paper is structured as follows.
%--------------------------------
In Secs.~\ref{Jpsi:etac:EM:form:factor} and
\ref{NRQCD:factorization:production:rate}, we express the product
rate for $e^+ e^- \to J/\psi + \eta_c$ in terms of the $J/\psi +
\eta_c$ electromagnetic (EM) form factor, and present the NRQCD
factorization formulas for both quantities, accurate through
relative order-$v^2$.
%--------------------------------
In Sec.~\ref{LO:result}, we list the tree-level short-distance
coefficients through relative ${\cal O}(v^2)$.
%--------------------------------
In Sec.~\ref{NLO:result}, we first sketch some key technical steps
about the NLO perturbative calculations, then present the asymptotic
expressions for the matching coefficients.
%--------------------------------
We devote Sec.~\ref{phenomenology} to exploring the phenomenological
impact of our new ${\mathcal O}(\alpha_s v^2)$ correction on the $B$
factory measurement.
%--------------------------------
Finally, we summarize in Sec.~\ref{summary}.
%--------------------------------

%----------------------------------------------------------------------
\section{$\bm{J}\bm{/}\bm{\psi}\bm{+}\bm{\eta}_{\bm c}$ EM form factor}
%--------------------------------
\label{Jpsi:etac:EM:form:factor}
%----------------------------------------------------------------------

Suppose we work in the $e^-$ and $e^+$ center-of-mass frame with
invariant mass of $\sqrt{s}$. Let $P_1$ ($\lambda$) denote the
momentum (helicity) of the $J/\psi$, and $P_2$ the momentum of the
$\eta_c$, respectively. This process simply probes the
$J/\psi+\eta_c$ electromagnetic form factor in the timelike region,
referred to as $G(s)$ hereafter:
%-----------------------------
\bqa
%-----------------------------
\langle J/\psi(P_1,\lambda)+\eta_c(P_2)\vert J_{\rm EM}^\mu \vert 0
\rangle = i\,G (s)\,\epsilon^{\mu\nu\rho\sigma} P_{1\nu} P_{2 \rho}
\varepsilon^*_\sigma (\lambda),
%-----------------------------
\label{EMFM:jpsi+etac}
%-----------------------------
\eqa
%-----------------------------
where $J^\mu_{\rm EM}$ is the electromagnetic current. The tensor
structure specified in (\ref{EMFM:jpsi+etac}) is uniquely dictated
by the Lorentz and parity invariance. As a result, the outgoing
$J/\psi$ must be transversely polarized, i.e., $\lambda=\pm 1$.

The cross section can be expressed as
%----------------------------
\bqa
%----------------------------
& & \sigma[e^+e^-\to J/\psi + \eta_c] = {4\pi \alpha^2 \over 3}
\left({|{\bf P}| \over \sqrt{s}}\right)^3 \left|G(s)\right|^2,
%----------------------------
\label{integrated:cross:section:Jpsi:etac}
%---------------------------------
\eqa
%----------------------------
where $|{\bf P}|$ signifies the magnitude of the 3-momentum carried
by the $J/\psi$ ($\eta_c$) in the center-of-mass frame. The cubic
power of $|{\bf P}|$ is reminiscent of the fact that $J/\psi$ and
$\eta_c$ are in the relative $P$-wave orbital state.

It is worth recalling the asymptotic behavior of
(\ref{integrated:cross:section:Jpsi:etac}) for this helicity-flipped
process. The helicity selection rule~\cite{Brodsky:1981kj} dictates
that $G(s)\sim 1/s^2$ as $\sqrt{s}\gg m_c$~\cite{Bondar:2004sv},
hence $\sigma[J/\psi+\eta_c] \sim 1/s^4$. Evidently, the charm quark
mass may serve as the agent of violating the hadron helicity
conservation. As will be examined in Sec.~\ref{NLO:result}, this
power-law scaling is subject to double-logarithmic modification once
beyond LO in $\alpha_s$.

%----------------------------------------------------------------------
\section{NRQCD factorization formula for $\bm{J}\bm{/}\bm{\psi}\bm{+}\bm{\eta}_{\bm c}$
production rate}
\label{NRQCD:factorization:production:rate}
%----------------------------------------------------------------------

According to NRQCD factorization formula, the $J/\psi+\eta_c$ EM
form factor in (\ref{EMFM:jpsi+etac}) can be factorized as
%---------------------------------
\bqa
%---------------------------------
\label{EMFF:NRQCD:factorization}
%---------------------------------
G(s) &=&  \sqrt{4 M_{J/\psi} M_{\eta_c}}\langle J/\psi| \psi^\dagger
\bm{\sigma} \cdot \bm{\epsilon} \chi |0\rangle \langle \eta_c|
\psi^\dagger \chi | 0 \rangle \nn\\
&&\left[c_0 + c_{2,1} \langle v^2 \rangle_{J/\psi} + c_{2,2} \langle
v^2 \rangle_{\eta_c} + \cdots \right],
%---------------------------------
\eqa
%---------------------------------
where $c_0$ and $c_2$ are the corresponding short-distance
coefficients. We have adopted relativistic normalization for the
quarkonia states appearing in the left side, while using the
nonrelativistic normalization for those in the NRQCD matrix
elements. For simplicity, we have introduced the following
dimensionless ratios of NRQCD matrix elements to characterize the
${\mathcal O}(v^2)$ corrections:
%------------------
%------------------
\begin{subequations}
%------------------
\bqa
%------------------
\langle v^2 \rangle_{J/\psi}  &=& {\langle J/\psi(\lambda)|
\psi^\dagger (-\tfrac{i}{2}\tensor{\mathbf{D}})^{2}
\bm{\sigma}\cdot\bm{\epsilon}(\lambda)\chi|0\rangle \over m_c^2\,
\langle J/\psi(\lambda)| \psi^\dagger
\bm{\sigma}\cdot\bm{\epsilon}(\lambda)\chi|0\rangle},
%------------------
\\
\langle v^2 \rangle_{\eta_c}  &=& {\langle \eta_c| \psi^\dagger
(-\tfrac{i}{2}\tensor{\mathbf{D}})^{2} \chi|0\rangle \over m_c^2\,
\langle \eta_c| \psi^\dagger \chi|0\rangle},
\label{v2:etac:definition}%
\eqa
%------------------
\end{subequations}
%------------------
where $\psi^\dagger \tensor{\mathbf{D}}\chi\equiv \psi^\dagger
{\mathbf{D}} \chi- ({\mathbf{D}} \psi)^\dagger \chi$.

Substituting Eq.~(\ref{EMFF:NRQCD:factorization}) into
(\ref{integrated:cross:section:Jpsi:etac}), one can decompose the
cross section into the ${\cal O}(v^0)$ and ${\cal O}(v^2)$ pieces:
%------------------
\bqa
%------------------
\label{cross:section:decomposition}
%------------------
& & \sigma [e^+e^-\to J/\psi+\eta_c] = \sigma_0+\sigma_2+{\mathcal
O}(\sigma_0 v^4),
%------------------
\eqa
%------------------
where
%------------------
\begin{subequations}
%------------------
\bqa
%------------------
\sigma_0 & = & {8\pi\alpha^2 m_c^2 (1-4 r)^{3/2}\over 3} \langle
{\mathcal O}_1 \rangle_{J/\psi} \langle{\mathcal
O}_1\rangle_{\eta_c}|c_{0}|^2,
%------------------
\\
%------------------
\sigma_2 & = & {4\pi \alpha^2 m_c^2 (1-4 r)^{3/2}\over 3} \langle
{\mathcal O}_1 \rangle_{J/\psi} \langle{\mathcal
O}_1\rangle_{\eta_c}
%------------------
\label{sigma_2:definition}
\\
%------------------
&& \Bigg\{
\bigg( {1-10r \over 1-4r} |c_{0}|^2
 + 4\,{\rm Re}[c_{0} c_{2,1}^*]
\bigg) \langle v^2\rangle_{J/\psi}
%------------------
\nn\\
%------------------
&& + \bigg( {1-10 r \over 1-4r} |c_{0}|^2 + 4\,{\rm Re}[c_{0}
c_{2,2}^*] \bigg ) \langle v^2\rangle_{\eta_c} \Bigg\}.\nn
%------------------
\eqa
%------------------
\label{sigma0:sigma2:definitions}
\end{subequations}
%------------------
In deriving (\ref{sigma0:sigma2:definitions}), we have employed the
Gremm-Kapustin relation~\cite{Gremm:1997dq} $M_H^2\approx
4m_c^2(1+\langle v^2 \rangle_{H})$ to eliminate the explicit
occurrences of $M_{J/\psi}$ and $M_{\eta_c}$. For notational
brevity, we have introduced the following symbols: $r={4 m_c^2 \over
s}$, $\langle \mathcal{O}_1\rangle_{J/\psi}=\big| \langle
J/\psi|\psi^\dag \bm{\sigma} \cdot \bm{\epsilon} \chi |0\rangle
\big|^2$, and $\langle {\mathcal O}_1 \rangle_{\eta_c} = \big|
\langle {\eta_c}| \psi^\dag \chi |0\rangle \big|^2$.

It is convenient to organize the short-distance coefficients $c_i$
in power series of the strong coupling constant, i.e.,
$c_i=c_i^{(0)}+{\alpha_s\over \pi}c_i^{(1)}+\cdots$. Accordingly, we
may decompose the cross section $\sigma_i$ into
$\sigma_i^{(0)}+\sigma_i^{(1)}$ ($i=0,2$). Our primary goal in this
work is to calculate $\sigma_2^{(1)}$.

%----------------------------------------------------------------------
\section{Tree-level NRQCD short-distance coefficients}
\label{LO:result}
%----------------------------------------------------------------------

\begin{figure}[tbH]
\begin{center}
\includegraphics[scale=0.43]{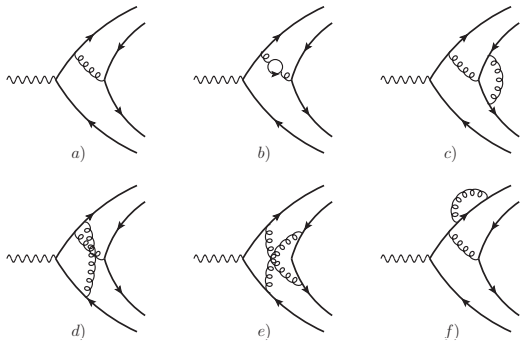}
\caption{One sample LO diagram and five sample NLO diagrams that
contribute to $\gamma^*\to J/\psi+\eta_{c}$.
\label{feynman:diagrams}}
\end{center}
\end{figure}

It is straightforward to employ the perturbative matching method to
determine the short-distance coefficients, by replacing the $J/\psi$
and $\eta_c$ states with the free $c\bar{c} (^3S_1^{(1)})$ and
$c\bar{c} (^1S_0^{(1)}) $ pairs, and enforcing that both
perturbative QCD and NRQCD calculations in
(\ref{EMFF:NRQCD:factorization}) yield the same answer, order by
order in $\alpha_s$.

There are in total 4 diagrams at LO in $\alpha_s$, one of which is
depicted in Fig.~\ref{feynman:diagrams}$(a)$. The tree-level
short-distance coefficients through ${\mathcal O}(v^2)$ have been
available long ago~\cite{Braaten:2002fi}. Here we list their values
at $D=4-2\epsilon$ spacetime dimension:
%------------------
\begin{subequations}
%------------------
\bqa
%------------------
c_{0}^{(0)}&=& {32 \pi C_F e_c \alpha_s \over N_c\,m_c s^2},
%------------------
\\
%------------------
c_{2,1}^{(0)}&=& c_{0}^{(0)} \left[\frac{3-10r}{6}  + \left(
1-\frac{16}{9}r \right)\epsilon+{\mathcal O}(\epsilon^2) \right],
%------------------
\\
%------------------
c_{2,2}^{(0)}&=& c_{0}^{(0)} \left[ \frac{2-5r}{3} + \left(
\frac{10}{9} - \frac{16}{9}r \right) \epsilon +{\mathcal
O}(\epsilon^2) \right],
%------------------
\eqa
%------------------
\label{c_i:tree-level:values}
\end{subequations}
%------------------
where $e_c={2\over 3}$ is the electric charge of the charm quark,
$N_c=3$ is the number of colors and $C_F={N_c^2-1\over 2N_c}$.

%%%%%%%%%%%%%%%%%%%%%%%%%%%%%%%%%%%%%%%%%%%%%%%%%%%%%%%%%%%%%%%%%%%%%%%%%%%%%%%%%%%%%%%%%%
%----------------------------------------------------------------------
\section{NLO perturbative NRQCD short-distance coefficients}
\label{NLO:result}
%----------------------------------------------------------------------

In this section, we first sketch some technical issues about the NLO
perturbative calculations, then present the analytic expressions of
the NLO corrections to the matching coefficients $c_0$ and $c_2$.
Technical details will be expounded in a long
write-up~\cite{Dong:2012:longpaper}.

%----------------------------------------------------------------------
\subsection{Description of the calculation}
%----------------------------------------------------------------------

The main technical difficulty is to identify the relativistic
effects in the quark amplitude $\gamma^*\to c\bar{c} (P_1,
^3S_1^{(1)})+ c\bar{c} (P_2, ^1S_0^{(1)})$ through NLO in
$\alpha_s$. The momenta of $c$ and $\bar{c}$ inside each pair are
assigned as
%------------------
$
%------------------
p_i = {1 \over 2}P_i+q_i, \ \bar{p}_i = {1 \over 2}P_i-q_i
%------------------
$
%------------------
for $i=1,2$. The total momentum $P_i$ and relative momentum $q_i$
are chosen to be orthogonal. In the rest frame of each $c\bar{c}$
pair, the total and relative 4-momenta read
$P_i^\mu=(2E_{q_i},\mathbf{0})$, $q_i^\mu=(0,{\mathbf q}_i)$,
respectively, where $E_{q_i}=\sqrt{m_c^2+{\mathbf q}_i^2}$.

We first employ the \textsc{Mathematica} package
\textsc{FeynArts}~\cite{Kublbeck:1990,Hahn:2000kx} to generate
Feynman diagrams and amplitudes for the process $\gamma^*\to
c(p_1)\bar{c}(\bar{p}_1)+c(p_2)\bar{c}(\bar{p}_2)$ to NLO in
$\alpha_s$. In total there are 20 two-point, 20 three-point, 18
four-point, and 6 five-point one-loop diagrams, some of which have
been illustrated in Fig.~\ref{feynman:diagrams}$(b)$-$(f)$. We then
apply the covariant spin projector~\cite{Bodwin:2002hg} to enforce
two $c\bar{c}$ pairs to form the spin-triplet/singlet, color-singlet
states, with the Dirac and color traces handled by
\textsc{FeynCalc}~\cite{Metig:1991}.

We then expand the amplitude in powers of the quark relative
momenta, $q_i$, up to the quadratic order. We then make the
following substitution to project out the $S$-wave states:
%------------------
\bqa
%------------------
q_i^\mu q_i^\nu  & \to & {{\mathbf q}_i^2 \over
D-1}\big(-g^{\mu\nu}+{P_i^\mu P_i^\nu\over P_i^2}\big),
%------------------
\eqa
%------------------
for $i=1,2$, and ${\mathbf q}_i^2$ is understood to be defined in
the rest frame of each $c\bar{c}$ pair.

In conventional matching procedure, one expands the relative
momentum $q_i$ only after completing the loop integrations in the
QCD amplitude, which is a daunting task in our case since the
entanglement of three disparate scales, $\sqrt{s}$, $m_c$, ${\bf
q}_i$, in a loop integral. In this work, we employ a much simpler
shortcut suggested by the {\it method of
region}~\cite{Beneke:1997zp}, i.e., making expansion in $q_i$ prior
to carrying out the loop integration. This amounts to directly
extracting the NRQCD short-distance coefficients, {\it i.e.}, the
contributions solely arising from the {\it hard} region ($k^2 \ge
m_c^2$). Consequently, we will no longer be distracted by the
effects from the low-energy regions such as the {\it potential}
($k^0\sim mv^2, |{\bf k}|\sim mv$) region.

To proceed, we use the \textsc{Mathematica} packages
\textsc{Fire}~\cite{Smirnov:2008iw} and \textsc{Apart}~\cite{apart}
to reduce the general higher-point one-loop tensor integrals into a
set of master integrals. As a bonus of having expanded the integrand
in powers of $q_i$, and utilized a trick of rescaling the pair
momentum $P_i$ to make some hidden relativistic effects explicit, it
turns out that all the required master integrals are just the usual
1-, 2- and 3-point scalar integrals, all of which can be found in
the Appendix of Ref.~\cite{Gong:2007db}.

We adopt the dimensional regularization to regularize both UV and IR
singularities, with spacetime dimension $D=4-2\epsilon$. We use 't
Hooft-Veltman scheme to handle
$\gamma_5$~\cite{'tHooft:1972fi,Breitenlohner:1977hr}. After summing
the contributions from all the diagrams, and incorporating mass and
coupling constant renormalization, we find that the ultimate NLO
expressions for the ${\mathcal O}(v^0)$ QCD amplitude are both UV
and IR finite, while the ${\mathcal O}(v^2)$ amplitude is UV finite
albeit IR divergent.

The occurrence of the IR divergences in the hard region at ${\cal
O}(\alpha_s v^2)$ is just as expected. From the pull-up mechanism,
one can identify the IR divergences encountered in the hard-region
calculation with those would arise from the {\it soft} ($k^\mu\sim
mv$) region in a literal QCD-side calculation, which must be
canceled out upon matching. In fact, these IR divergences can be
reconstructed with the knowledge of the ${\cal O}(\alpha_s v^2)$
correction to the perturbative NRQCD matrix elements $\langle
c\bar{c}(^3S_1)| \psi^\dag \bm{\sigma} \cdot \bm{\epsilon} \chi
|0\rangle^{(1)}$~\cite{Luke:1997ys}, $\langle c\bar{c}(^1S_0)|
\psi^\dag \chi |0\rangle^{(1)}$~\cite{Jia:2011ah}, as well as
$c_0^{(0)}$~\cite{Dong:2012:longpaper}. Thus, we finally end up with
the both UV, IR-finite short-distance coefficients $c_i^{(1)}$
($i=0,2$). Our finding is consistent with the
all-order-in-$\alpha_s$ proof outlined in \cite{Bodwin:2008nf}, that
NRQCD factorization holds for the exclusive production of one
$S$-wave quarkonium plus any higher orbital angular momentum
quarkonium in $e^+e^-$ annihilation.

%----------------------------------------------------------------------
\subsection{Asymptotic expressions of NLO short-distance coefficients}
%----------------------------------------------------------------------

The NRQCD short-distance coefficients $c_0^{(1)}$ and $c_2^{(1)}$
are in general complex-valued. Their analytic expressions are
somewhat lengthy and will not be reproduced here. Nevertheless, it
is enlightening to know their asymptotic expressions in the limit
$\sqrt{s}\gg m_c$:
\begin{widetext}
%-------------------
\allowdisplaybreaks
\begin{subequations}
%-------------------
\bqa
%-------------------
c_{0}^{(1)}\left( r,{\mu_r^2 \over s} \right)_{\rm asym} &=&
c_{0}^{(0)}\times \Bigg\{ \beta_0 \bigg(- \frac{1}{4}
\ln\frac{s}{4\mu_r^2} + \frac{5}{12}\bigg) + \bigg( \frac{13}{24}
\ln^2 r + \frac{5}{4} \ln2 \ln r - \frac{41}{24} \ln r -
\frac{53}{24}\ln^22 +\frac{65}{8}\ln2
%-------------------
\nn\\
%-------------------
&& -\frac{1}{36}\pi^2-\frac{19}{4} \bigg)
 + i\pi \bigg( \frac{1}{4}\beta_0 + \frac{13}{12}\ln r  +
\frac{5}{4} \ln2- \frac{41}{24} \bigg)  \Bigg\},
%-------------------
\label{c_0:NLO:asym:expressions}
\\
%-------------------
c_{2,1}^{(1)}\left( r,{\mu_r^2 \over s},{\mu_f^2 \over m_c^2}
\right)_{\rm asym} &=& {1\over 2} c_{0}^{(0)}\times \Bigg\{
\frac{16}{9}\ln\frac{\mu_f^2}{m_c^2} + \beta_0 \bigg( -\frac{1}{4}
\ln\frac{s}{4\mu_r^2} + \frac{11}{12}\bigg) + \bigg( \frac{3}{8}
\ln^2 r + \frac{19}{12} \ln2 \ln r + \frac{31}{24} \ln r -
\frac{1}{24}\ln^22
%-------------------
\nn\\
%-------------------
&& +\frac{893}{216}\ln2-\frac{5}{36}\pi^2-\frac{497}{72} \bigg)
 + i\pi \bigg( \frac{1}{4}\beta_0 + \frac{3}{4}\ln r +
\frac{19}{12} \ln2 + \frac{9}{8} \bigg)  \Bigg\},
\\
%-------------------
c_{2,2}^{(1)}\left( r,{\mu_r^2 \over s},{\mu_f^2 \over m_c^2}
\right)_{\rm asym} &=&  {2\over 3} c_{0}^{(0)}\times  \Bigg\{
\frac{4}{3}\ln\frac{\mu_f^2}{m_c^2} +  \beta_0 \bigg( -\frac{1}{4}
\ln\frac{s}{4\mu_r^2} + \frac{2}{3}\bigg) + \bigg( \frac{1}{12}
\ln^2 r + \frac{11}{12} \ln2 \ln r  - \frac{1}{24} \ln r -
\frac{11}{8}\ln^22
%-------------------
\nn\\
%-------------------
&& +\frac{241}{144}\ln2-\frac{1}{8}\pi^2 - \frac{99}{16} \bigg)
 + i\pi \bigg( \frac{1}{4} \beta_0 + \frac{1}{6}\ln r +
\frac{11}{12} \ln2 - \frac{1}{24} \bigg)  \Bigg\},
%-------------------
\eqa
%-------------------
\label{c_i:NLO:asym:expressions}
\end{subequations}
%-------------------
\end{widetext}
where $\beta_0={11\over 3} C_A-{2\over 3}n_f$ is the one-loop
coefficient of the QCD $\beta$ function, and $n_f=4$ denotes the
number of active quark flavors. $\mu_r$ denotes the renormalization
scale, with the natural magnitude of order $\sqrt{s}$; and $\mu_f$
is identified with the factorization scale in the $\overline{\rm
MS}$ scheme, whose value lies in somewhere between $m_c v$ and
$m_c$, the UV cutoff scale of NRQCD.

It has been pointed out~\cite{Jia:2010fw} that a peculiar
double-logarithmic correction $\propto \ln^2 r$ arises in
$c_0^{(1)}$ for this helicity-flipped process, and our
(\ref{c_0:NLO:asym:expressions}) exactly agrees with the
corresponding expression there.
Equations~(\ref{c_i:NLO:asym:expressions}) imply that the presence
of the double logarithm persists to ${\cal O}(v^2)$ as well.
Presumably, provided that $\sqrt{s}\gg m_c$, one must resum these
types of large logarithms to all orders to obtain a reliable
prediction. Unfortunately, due to our incapability for the
light-cone approach to dealing with the helicity-suppressed hard
exclusive processes beyond LO in $\alpha_s$, currently it remains to
be an open question how to fulfill such a resummation.

%----------------------------------------------------------------------
\section{Phenomenology}
\label{phenomenology}
%----------------------------------------------------------------------

In the numerical analysis, we take $\sqrt{s}=10.58$ GeV, and the QED
coupling constant $\alpha(\sqrt{s}) = 1/130.9$~\cite{Bodwin:2007ga}.
The running strong coupling constant is evaluated by using the
two-loop formula with $\Lambda^{(4)}_{\overline{\rm MS}}= 0.338$
GeV~\cite{Zhang:2005cha,Gong:2007db}. The LO NRQCD matrix elements
are taken from~\cite{Dong:2011fb}: $ \langle
{\mathcal{O}_1}\rangle_{J/\psi}\approx\langle
{\mathcal{O}_1}\rangle_{\eta_c}=0.387\;{\rm GeV}^3 $. If we choose
$m_c=1.4$ GeV ($r=0.0700$), the Gremm-Kapustin relation implies that
$\langle v^2\rangle_{J/\psi} = 0.223$ and $\langle
v^2\rangle_{\eta_c} = 0.133$.  We will take $\mu_f=m_c$.

We are ready to carry out a detailed analysis for the processes
$e^+e^-\to J/\psi + \eta_c$ and confront the $B$ factory
measurements. One important source of theoretical uncertainties
comes from the scale setting for the strong coupling constant. There
is no way to circumvent the scale ambiguity problem within the
confine of NRQCD factorization, and we proceed to estimate the cross
section by affiliating all the occurring $\alpha_s$ with a common
scale, $\mu_r$, and choosing $\mu_r=\sqrt{s}/2$ and $\mu_r=2 m_c$,
respectively. It is hoped that the less biased results interpolate
between these two sets of predictions.

With $r=0.07$, for $\mu_r=\sqrt{s}/2$, we find $c_0^{(1)}/c_0^{(0)}
= 8.83-6.02i$, $c_{2,1}^{(1)}/c_{2,1}^{(0)} = -5.09+8.89 i$, and
$c_{2,2}^{(1)}/c_{2,2}^{(0)} = -5.15+8.99 i$; for $\mu_r=2m_c$ we
have $c_0^{(1)}/c_0^{(0)} = 6.18-6.02i$,
$c_{2,1}^{(1)}/c_{2,1}^{(0)} = -7.74 + 8.89 i$, and
$c_{2,2}^{(1)}/c_{2,2}^{(0)} = -7.80 + 8.99 i$. Since ${\rm
Re}[c_{2,i}^{(1)}/c_{2,i}^{(0)}]$ is large and negative for $i=1,
2$, in conjunction with (\ref{c_i:tree-level:values}), one may think
that the new ${\cal O}(\alpha_s v^2)$ correction would dilute the
known ${\cal O}(v^2)$ effect.

%--------------------------------------
\renewcommand\arraystretch{1.5}
\begin{table}[tbH]
\caption{Individual contributions to the predicted $\sigma[e^+e^-\to
J/\psi+\eta_{c}]$ at $\sqrt{s}=10.58$ GeV, labeled by powers of
$\alpha_s$ and $v$. The cross sections are in units of fb.
\hfill\hfill \label{tbl:cross:component}}
\begin{tabular}%{|c|c|c|c|c|}
{|>{\centering}p{0.15\textwidth}|>{\centering}p{0.07\textwidth}|>{\centering}p{0.07\textwidth}
|>{\centering}p{0.07\textwidth}|>{\hfill}p{0.07\textwidth}<{\hfill\hfill}|}
\hlinew{1pt}
   $\alpha_s(\mu_r)$  &  $\sigma_0^{(0)}$  & $\sigma_0^{(1)}$ & $\sigma_2^{(0)}$& $\sigma_2^{(1)}$ \\
\hline
   $\alpha_s(\frac{\sqrt{s}}{2}) = 0.211$  &  $4.40$  & $5.22$  & $1.72$ & $0.73$ \\
\hline
   $\alpha_s(2 m_c) = 0.267$  &  $7.00$  & $7.34$  & $2.73$ & $0.24$ \\
\hlinew{1pt}
\end{tabular}
\end{table}

\begin{figure}[tbH]
\begin{center}
\includegraphics[width=0.38\textwidth]{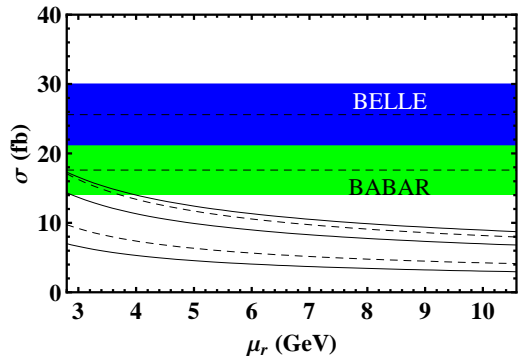}
\caption{The $\mu_r$ dependence of the cross section for $e^+e^-\to
J/\psi+\eta_{c}$ at $\sqrt{s}=10.58$ GeV. The 5 curves from bottom
to top are $\sigma_0^{(0)}$ (solid line),
$\sigma_0^{(0)}+\sigma_2^{(0)}$ (dashed line),
$\sigma_0^{(0)}+\sigma_0^{(1)}$ (solid line),
$\sigma_0^{(0)}+\sigma_2^{(0)}+\sigma_0^{(1)}$ (dashed line), and
$\sigma_0^{(0)}+\sigma_2^{(0)}+\sigma_0^{(1)}+\sigma_2^{(1)}$ (solid
line), respectively. The blue and green bands represent the measured
cross sections by the \textsc{Belle} and \textsc{BaBar} experiments,
with respective systematic and statistical errors added in
quadrature. \label{crossX:figure} }
\end{center}
\end{figure}

Table~\ref{tbl:cross:component} lists the predicted cross sections
for $e^+e^-\to J/\psi+\eta_{c}$ in double expansions of $\alpha_s$
and $v$, with two sets of $\mu_r$. We reproduce the well-known
results, i.e., the positive and significant ${\mathcal O}(\alpha_s)$
correction~\cite{Zhang:2005cha,Gong:2007db}, and the positive but
less pronounced ${\mathcal O}(v^2)$
correction~\cite{Braaten:2002fi,He:2007te,Bodwin:2007ga}. The new
${\cal O}(\alpha_s v^2)$ ingredient, $\sigma_2^{(1)}$, is positive
but modest. It may be attributed to the {\it near cancelation}
between the two terms in the prefactor of $\langle v^2 \rangle_{H}$
in (\ref{sigma_2:definition}).

In Fig.~\ref{crossX:figure}, we plot the $\sigma[e^+e^-\to
J/\psi+\eta_{c}]$ as a function of $\mu_r$. Whether including the
contribution of $\sigma_2^{(1)}$ or not clearly does not make a big
difference. When $\mu_r$ is relatively small, the state-of-the-art
NRQCD prediction converges to the \textsc{Babar} measurement within
errors. Had we taken somewhat larger values of the NRQCD matrix
elements $\langle {\mathcal{O}_1}\rangle_{H}$ as in
\cite{He:2007te,Bodwin:2007ga}, the agreement with the two $B$
factories measurements would be better.

%----------------------------------------------------------------------
\section{Summary}
\label{summary}
%----------------------------------------------------------------------

In this work we have computed the ${\mathcal O}(\alpha_s v^2)$
correction to the helicity-suppressed process $e^+e^-\to
J/\psi+\eta_c$ in the NRQCD factorization framework. The
corresponding NLO perturbative short-distance coefficients
associated with the $J/\psi+\eta_c$ EM form factor are directly
extracted from the {\it hard} loop-momentum region through relative
order $v^2$. By examining the asymptotic form of these coefficients,
we reconfirm the pattern recognized in
\cite{Jia:2010fw,Dong:2011fb}: The hard exclusive processes
involving double-charmonium at higher twist in general are plagued
with double logarithms of form $\ln^2{s/m_c^2}$ once beyond LO in
$\alpha_s$. When $\sqrt{s}\gg m_c$, in order to obtain the reliable
predictions for such types of processes, one is enforced to resum
these potentially large double logarithms to all orders in
$\alpha_s$, which so far remains to be an open challenge.

At the $B$ factory energy, we found that incorporating this new
piece of correction only modestly enhances the existing NRQCD
predictions. This may be ascribed to some accidental cancelation
between two different sources of relativistic corrections. At much
higher $\sqrt{s}$, this ${\mathcal O}(\alpha_s v^2)$ correction
would be much more relevant.

%--------------------------------------------------------------------
\begin{acknowledgments}
%--------------------------------------------------------------------
We thank Wen-Long Sang for valuable discussions.
%--------------------------------------------------------------------
This research was supported in part by the National Natural Science
Foundation of China under Grant No.~10875130, No.~10935012, and by
China Postdoctoral Science Foundation.
%--------------------------------------------------------------------
\end{acknowledgments}
%--------------------------------------------------------------------

\end{document}